\documentclass{article}

\usepackage{arxiv}

\usepackage[utf8]{inputenc} % allow utf-8 input
\usepackage[T1]{fontenc}    % use 8-bit T1 fonts
\usepackage{hyperref}       % hyperlinks
\usepackage{url}            % simple URL typesetting
\usepackage{booktabs}       % professional-quality tables
\usepackage{amsfonts}       % blackboard math symbols
\usepackage{nicefrac}       % compact symbols for 1/2, etc.
\usepackage{microtype}      % microtypography
\usepackage{lipsum}
\usepackage{setspace}
\usepackage{lineno}
\usepackage{amsmath}
\usepackage{graphicx}
\usepackage{float}

\title{Variational Analysis of Landscape Elevation and Drainage Networks}

\author{
  Milad Hooshyar \\
  CEE, PEI, and PIIRS\\
  Princeton University\\
  \texttt{hooshyar@princeton.edu}\\
   \And
 Shashank Kumar Anand \\
  CEE\\
  Princeton University\\
  \texttt{skanand@princeton.edu} \\
   \And
 Amilcare Porporato \thanks{Corresponding author}\\
  CEE and PEI\\
  Princeton University\\
  \texttt{aporpora@princeton.edu} \\
  }

\begin{document}
\maketitle

\begin{abstract}
Landscapes evolve toward surfaces with complex networks of channels and ridges in response to climatic and tectonic forcing. Here we analyze variational principles giving rise to  minimalist models of landscape evolution as a system of partial differential equations that capture the essential dynamics of sediment and water balances. Our results show that in the absence of diffusive soil transport, the steady-state surface extremizes the average domain elevation. Depending on the exponent $m$ of specific drainage area in the erosion term, the critical surfaces are either minima ($0<m<1$) or maxima ($m>1$), with $m=1$ corresponding to a saddle point. Our results establish a connection between Landscape Evolution Models (LEMs) and Optimal Channel Networks (OCNs) and elucidate the role of diffusion in the governing variational principles.   
\end{abstract}

% keywords can be removed

\section{Introduction}
The dynamic of the land surface elevation is governed by the balance between soil erosion, deposition, and production, which are modulated by climate and tectonic activities \cite{bonetti2019Cascade, sweeney2015experimental, perron2009formation, izumi1995inception}. When soil erosion is driven by the water movement on the surface, landscape evolution gives rise to networks of channels and ridges with scaling laws typical of complex systems \cite{horton1932drainage, shreve1974variation, tarboton1988fractal, Rodriguez2001}. Minimalist mathematical models and physical experiments produce surface patterns resembling those observed in nature, helping to shed light on their statistical and scaling behavior under different conditions of external forcing \cite{sweeney2015experimental, singh2015landscape, perron2008controls, bonetti2019Cascade}.   

The landscape elevation is linked to the surface sediment budget, accounting for soil creep, erosion, and uplift \cite{izumi1995inception, howard1994detachment,fowler2011mathematical, smith2010theory}. The erosion at a point is commonly assumed to be a power function of slope and water flux; thus, the sediment budget equation is coupled to an equation for the specific contributing area, resulting from a simplified water continuity equation \cite{bonetti2018theory}. This minimalist model produces a  channelization sequence, the intensity of which is modulated by the relative magnitude of erosive to diffusive transport, with intriguing analogies with other complex non-equilibrium systems such as fluid turbulence \cite{bonetti2019Cascade,hooshyar_log}. 

\begin{figure}[H]
  \centering
  \includegraphics[width=0.7\textwidth]{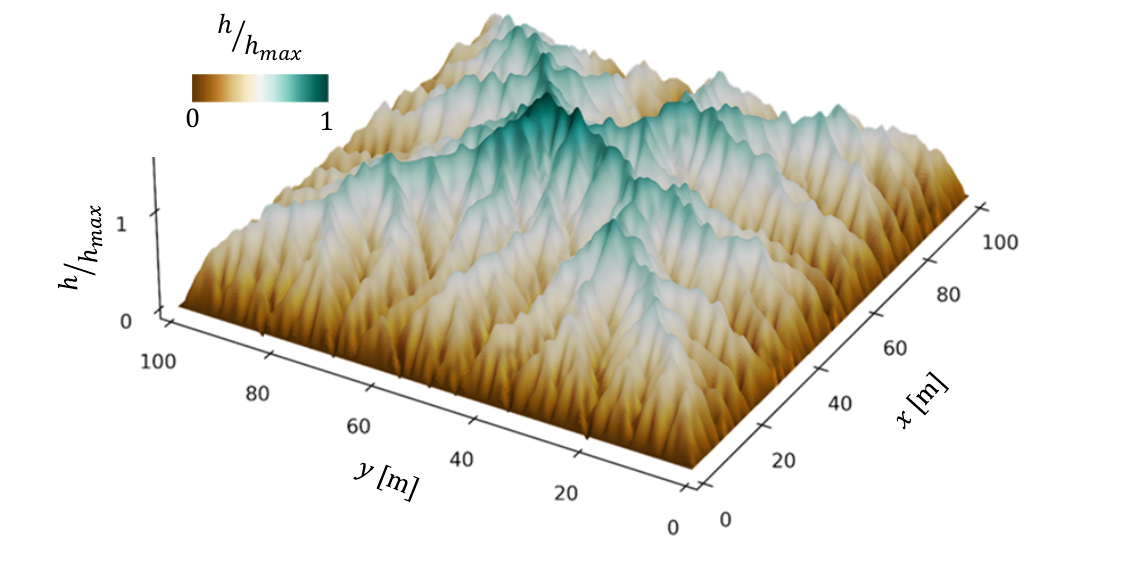}
  \caption{The channelized surface generated by numerical solution of Eqs. \eqref{eq:lem} and \eqref{eq:water_depth}, when erosion is dominant over diffusive soil transport. The surface elevation is denoted by $h$ and $h_{max}$ is the maximum elevation. Model parameter in Eq. \eqref{eq:lem} are $m = 0.5$, $D=0.005\ \text{m}^2 \text{y}^{-1}$, $K=\ \text{m}^{0.5} \text{y}^{-1}$, and $U=0.01\ \text{m} \text{y}^{-1}$. }
  
  \label{fig:3D}
\end{figure}

The intertwined channel and ridge networks that emerge from this type of evolution (see Fig. \ref{fig:3D} for example)  are suggestive of an optimal transport strategy as observed in the branched regime of Monge-Kantorovich problem where the cost function favors mass aggregation \cite{bergamaschi2019spectral, santambrogio2015optimal,brasco2011benamou}. The optimality principle behind river networks was first studied within the context of so-called Optimal Channel Networks (OCNs) and, more generally in the context of optimal transportation theory \cite{balister2018river, rinaldo2014evolution, banavar1997sculpting, Rigon1993}. OCNs are the configurations that connect sites over a discretized domain while minimizing the total energy loss of the water flow \cite{rinaldo1996thermodynamics, balister2018river}. Depending on the exponent of the drainage area used for the computation of local energy dissipation (denoted by $\gamma$), the optimal configurations range from spanning trees, resembling natural river networks for $0<\gamma <1$ \cite{rinaldo1996thermodynamics}, to spiral networks, for $\gamma < 0$ \cite{banavar2000topology}.

The connection between OCNs and Landscape Evolution Models (LEMs) has been investigated in several contributions \cite{sinclair1996mechanism, banavar2001scaling, rinaldo2014evolution, balister2018river}. Most significantly, it has been shown that OCNs are stationary solutions of a sediment transport law. They correspond to landscapes which accommodate flow toward the steepest slope\cite{banavar2001scaling, balister2018river}. Such results are with strong discretization \cite{santambrogio2007optimal} in which the surface is approximated by a network of connected nodes with links to represent the hydraulic connections. The generalization of these findings and the connection to model in the continuous domain has received less attention in the literature.  

In this paper we show that the steady-state landscapes obtained by solving the governing PDEs of sediment and water continuity equations in a continuous domain and under negligible soil diffusion assumption, reach an optimal state, defined based on the average domain elevation. We prove that each of the two governing PDEs can be replaced by an corresponding variational principle. By investigating the second variation, we show that the optimal states may be reached either by minimizing or maximizing the average elevation based on the exponent of the specific drainage area in the erosion law. The role of the diffusion on the underlying variational principle is also discussed.  

\section{Landscape evolution models (LEMs)}\label{se: LEM}

In LEMs, the land surface elevation $h$ evolves as a function of diffusive soil creep, erosion, deposition, and tectonic uplift. When the deposition of mobilized sediment is negligible (detachment-limited condition), the surface evolution can be written according to \cite{howard1994detachment, izumi1995inception}, 

\begin{equation}
\label{eq:lem}
\frac{\partial h}{\partial t}=D\Delta h-K a^m |\nabla h| + U, 
\end{equation}
where $a$ is the specific catchment area (or catchment area per unit contour length). The diffusive soil creep is given as $D\Delta h$ and $D$ is  the soil diffusivity. The term $K a^m |\nabla h|$ quantifies the  erosion (sediment movement due to water flow) with constants $m$ and $K$. The tectonic uplift rate is denoted by $U$. Under the assumption that water moves with constant speed in the direction of the steepest slope \cite{rodriguez1992energy},  the specific catchment area $a$ is proportional to the water height generated by a steady and spatially uniform rainfall, and is given by the steady-state continuity equation of water flow over the surface \cite{bonetti2018theory, bonetti2019Cascade}, 

\begin{equation}
\label{eq:water_depth}
\nabla.\left(a\frac{\nabla z}{|\nabla z|}\right) = -1.
\end{equation}

\begin{figure}[H]
  \centering
  \includegraphics[width=0.7\textwidth]{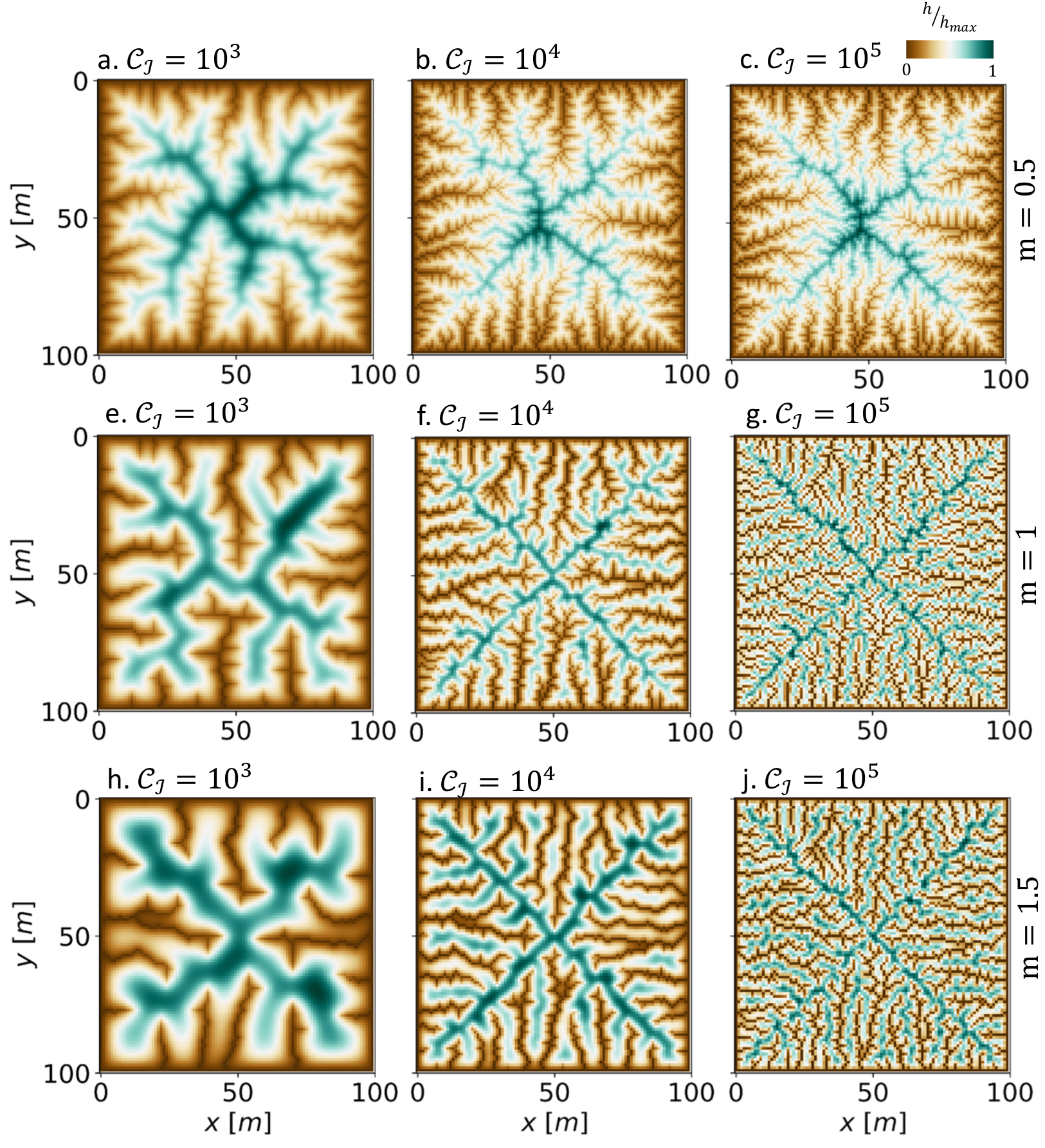}
  \caption{Numerical solution of the system of Eqs. \eqref{eq:lem} and \eqref{eq:water_depth} in a square domain ($100 \text{m}$ by $100 \text{m}$ with $1 \text{m}$ grid spacing) for $m=0.5$, $m=1$, and $m=1.5$ with increasing $C_I$. As the channelization index increases (higher erosion), the surface becomes more dissected with branching channels. The 3D view (b) is shown in Fig. \ref{fig:3D}. The numerical scheme uses an implicit solver for the erosion term where the specific drainage area $a$ is approximated by the $D_{\infty}$ flow direction method \cite{tarboton1997new, bonetti2019Cascade}. The model parameters are  $D=0.005\ \text{m}^2 \text{y}^{-1}$ and $U=0.01\ \text{m} \text{y}^{-1}$, where $K$ was appropriately selected to get the desired $\mathcal{C_I}$ for a given $m$.}
  \label{fig:numerical_ci}
\end{figure}

The coupled Eqs. \eqref{eq:lem} and \eqref{eq:water_depth} form a closed system. In a domain with a single length scale $l$ (e.g., a square or a semi-infinite stripe of constant width) the behavior of the system is controlled by the "channelization index" $\mathcal{C_I} = \frac{K l^{m+1}}{D}$ \cite{bonetti2018theory}. For low values of $\mathcal{C_I}$, the steady-state surfaces are smooth with no concentrated flow paths. As $\mathcal{C_I}$ exceeds a critical threshold, the surface becomes channelized with only first order non-branched channels. In the infinite stripe case, this threshold is $\mathcal{C_I}_{cr} \approx 37$ for $m=1$, a result derived by linearizing the system of equations around its unchanneled solution for $m=1$ \cite{bonetti2019Cascade}. Numerical simulations further showed that higher $m$ values correspond to smaller $\mathcal{C_I}_{cr}$ \cite{bonetti2019Cascade}. As the value of  $\mathcal{C_I}$ increases beyond $\mathcal{C_I}_{cr}$, the surface becomes progressively channelized with more branching. Fig. \ref{fig:numerical_ci} shows the controls of  $\mathcal{C_I}$ in the level of surface channelization for a set of numerical simulations in a $100\ \text{m}$ by $100\ \text{m}$ domain. The parameter $m$ also controls the structure of the channel networks: surfaces with higher $m$ values tend to have more branched networks with wider channels and higher bifurcation angles \cite{bonetti2019Cascade}.

\section{Optimal Channel Networks (OCNs)} \label{se:OCN}

As mentioned in the introduction, scaling laws of channel networks have been mostly studied in relations to the Optimal Channel Networks (OCNs). These are the configurations that are defined over a lattice of $N$ nodes and locally minimize  \cite{Rigon1993, rinaldo1996thermodynamics, rinaldo2014evolution}, 

\begin{equation}
\label{eq:OCN_obj}
J_{OCN}[c] = \sum_i A_i^{\gamma}, 
\end{equation}
where $\gamma$ is a constant and $A_i$ is the drainage area at node $i$ (total area which drains into node $i$ ),  uniquely defined based the connectivity matrix of the configuration $c$.  The previous condition is obtained by assuming a power relation between local slope and drainage area at each link (i.e., $|\nabla h|_i \propto A_i^{\gamma -1}$), so that functional \eqref{eq:OCN_obj} is proportional to the total dissipated energy of water flow. From the same power relation between local slope and drainage area, \cite{rinaldo1993self} showed that functional \eqref{eq:OCN_obj} is proportional to the average domain elevation of the discretized domain. 

OCNs give rise to several statistical scaling laws (e.g. Horton's laws of stream length and number \cite{horton1945erosional}) observed in the natural river networks \cite{Rodriguez2001}.  OCNs are commonly generated by numerical optimization of functional \eqref{eq:OCN_obj} over a lattice. Starting from an initial loopless random configuration spanning the whole domain, at each iteration, a random change in the configuration is tried. If this change results in a loopless configuration, the drainage area is computed from the altered connectivity matrix, which in turn, gives a new value of functional \eqref{eq:OCN_obj}. If the change reduces the numerical value of functional \eqref{eq:OCN_obj}, it is accepted; otherwise, the change is reversed. This process is repeated until a given number of iterations ($1000$ in this study) with no accepted changes are tried. This approach is referred to as greedy, in the sense that it only accepts or rejects random changes based on the immediate response of the objective functional; therefore, it tends toward a so-called "imperfect OCN" \cite{rinaldo2014evolution}. Imperfect OCNs are feasible (dynamically accessible) optimal state of the system given the initial condition and are shown to bear more resemblance to natural networks comparing to the global (or near-global) optimal configuration achieved by more sophisticated optimization algorithms such as simulated annealing \cite{kirkpatrick1983optimization, rinaldo1996thermodynamics, rinaldo2014evolution}.

\begin{figure}[H]
  \centering
  \includegraphics[width=0.7\textwidth]{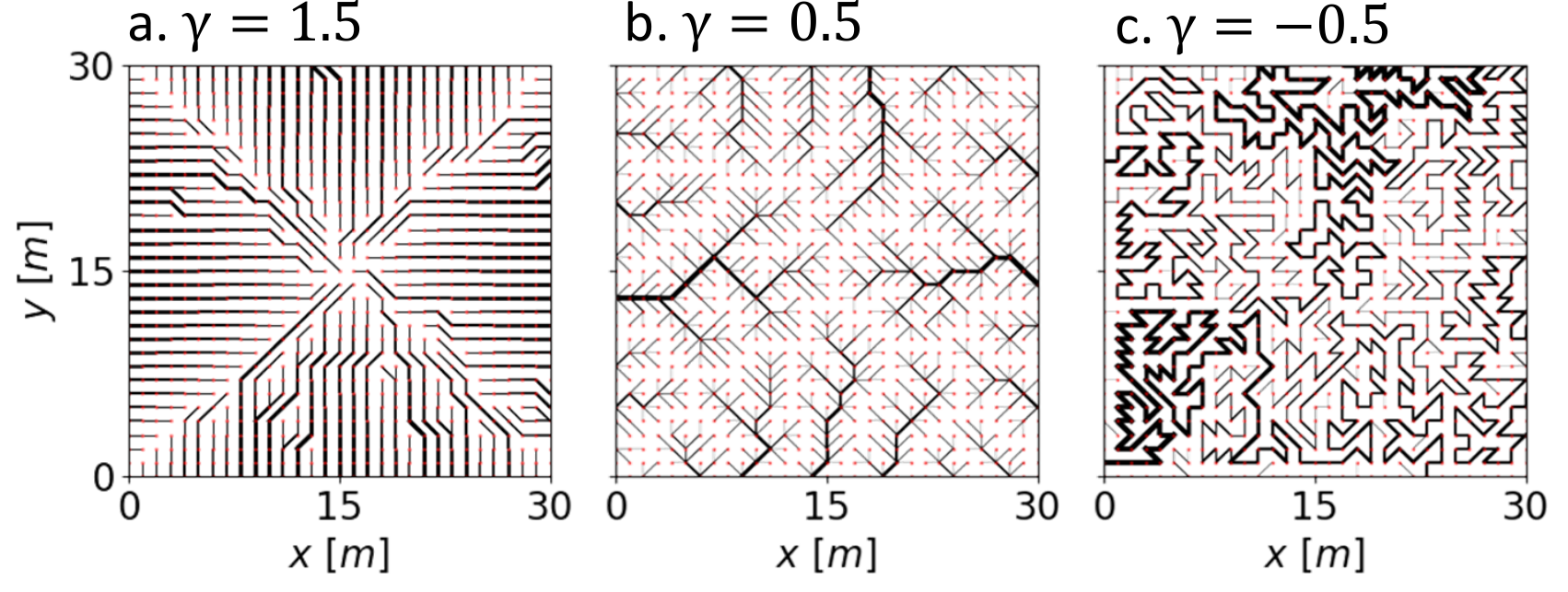}
  \caption{The configurations that locally minimize functional \eqref{eq:OCN_obj} in a $30\ \text{m}$ by $30$ $\text{m}$ domain with $1\ \text{m}$ grid spacing. (a) For $\gamma > 1$, the optimal configuration follows the shortest path to the outlet (domain boundary). (b) For $0<\gamma<1$, tree-like structures emerge as the optimal configuration which resemble natural river networks. (c) If only loop-less configurations are allowed, $\gamma < 0$ results in spiral configurations. These results are achieved by a greedy optimization algorithm starting from a random loop-less initial configuration. The width on the links are proportional to $\sqrt{A}$.}
  \label{fig:OCN}
\end{figure}

Fig. \ref{fig:OCN} shows the configurations corresponding to a local minimum of functional \eqref{eq:OCN_obj} for different values of $\gamma$. The range $0<\gamma<1$ leads to branching channels similar to natural rivers \cite{rinaldo1996thermodynamics}. For $\gamma > 1$, the  configuration minimizing energy expenditure connects each node to the outlet (domain boundaries in this case) following the shortest path  \cite{rinaldo1996thermodynamics}. For $\gamma < 0$, loops are preferred and if one excludes loops, spiral networks emerge \cite{banavar2000topology} with configurations very different from those observed in the natural river networks. The role of $\gamma$ is reminiscent of the behavior of the Monge-Kantorovich optimal transport, where congested (similar to Fig. \ref{fig:OCN}a) or branched (similar to Fig. \ref{fig:OCN}b) transport may occur depending on the exponent of the Wasserstein distance used to penalize the transport of mass \cite{brasco2011benamou, santambrogio2015optimal}. The extended dynamic Monge-Kantorovich formulation proposed by \cite{facca2018extended} also exhibits similar transition from congested to branched transport depending on the sub- or super-linear growth of the transport density with respect to the transport flux.

\section{Variational principle for LEMs in the absence of diffusion}

\subsection{The first variation}

We first  consider the case of negligible diffusive transport (i.e., $D=0$ or $\mathcal{C_I} \rightarrow \infty$). For this case, we will show that the steady-state solutions of Eqs. \eqref{eq:lem} and \eqref{eq:water_depth} with zero elevation at the boundaries ($h(x \in \Omega) = 0$) correspond to the  critical functions of the objective functional

\begin{equation}
\label{eq:functional_D0}
J[h] = \int_S h  ds, 
\end{equation}
where the integral is over the domain $S$ and $\Omega$ is the domain boundary. In particular, we prove in this section that the first variation of functional $\eqref{eq:functional_D0}$,  subject to the constraint of the sediment continuity equation (Eq. \eqref{eq:lem} at the steady-sate with $D=0$), vanishes when the $h$ and $a$ fields satisfy Eq. \eqref{eq:water_depth}. Functional \eqref{eq:functional_D0} is proportional to the average domain elevation and is equivalent to the mean potential energy of the surface water which gradually dissipates as it flows toward the domain boundaries.   

We first consider functional \eqref{eq:functional_D0} for the landscape elevation. Following the classical approach of the calculus of variation originally developed by Lagrange \cite{lanczos2012variational}, the  constraint of the sediment balance equation can be imposed using a Lagrange multiplier field $\lambda$ as 

\begin{equation}
\label{eq:functional_D0_C}
J[ h] = \int_S h + \lambda \left(-K a^m |\nabla h| + U \right) ds. 
\end{equation}
In this manner we ignore the specific area equation, which is replaced by the condition of extremization of functional \eqref{eq:functional_D0}, and thus treat $a$ independent of $h$. 

Introducing a variation $g$ of $h$ over the domain $S$ with the boundary conditions $g(x \in \Omega) = 0$, the first variation of the functional $J$ in Eq. \eqref{eq:functional_D0_C} is 

\begin{subequations}
\begin{align}
\label{eq:1st_var_1_D0}
\delta J[h, g] = \frac{dJ[h + \epsilon g]}{d\epsilon}\Bigg |_{\epsilon =0}  =  & \frac{d}{d \epsilon} \int_S (h + \epsilon g) + \lambda \left(-K a^m |\nabla h  +  \epsilon \nabla g| + U \right) ds\Bigg |_{\epsilon =0}  \nonumber \\
& =\int_S  g ds  + \int_S \lambda \left(-K a^m \frac{d}{d \epsilon}|\nabla h  +  \epsilon \nabla g| + U \right) ds\Bigg |_{\epsilon =0} \nonumber \\
 &=\int_S  g ds  - \int_S K  \lambda a^m  \frac{\nabla h}{|\nabla h|} \cdot \nabla g  ds. 
\end{align}
\end{subequations}

Integrating the second term by parts and using the boundary condition  $g(x \in \Omega) = 0$ leads to  

\begin{equation}
\label{eq:1st_var_2_D0}
\delta J[h, g]  = \int_S  g \left[1  - K  \nabla \cdot  \left(\lambda a^m  \frac{\nabla h}{|\nabla h|}\right)\right]  ds.
\end{equation}

For the critical functions, the first variation must vanish ($\delta J[h, g] = 0$) for any $g$; thus the fundamental lemma of calculus of variations \cite{giaquinta2013calculus1, lanczos2012variational} yields 

\begin{equation}
\label{eq:1st_var_3_D0}
1  - K  \nabla \cdot  \left(\lambda a^m  \frac{\nabla h}{|\nabla h|}\right) = 0.
\end{equation}

This shows that Eq. (\ref{eq:1st_var_3_D0}) is equivalent to Eq. \eqref{eq:water_depth} for $\lambda = \frac{1}{K}a^{1-m}$, proving our claim at the beginning of this section. Thus, the $a$ and $h$ fields that extremize functional \eqref{eq:functional_D0} and follow the erosion law given by Eq. \eqref{eq:water_depth} with $D=0$ and $\frac{\partial h}{\partial t}=0$, also satisfy Eq. \eqref{eq:water_depth} and vice-versa. This means that Eq. \eqref{eq:water_depth} in the system of equations governing landscape evolution can be replaced by a variational principle based on the mean elevation over the domain, that is Eq. \eqref{eq:functional_D0}.  

\subsection{Dual formulation}

It is interesting to show that the problem has a dual formulation in which the sediment continuity can be replaced by a variational principle based on a functional, constructed with the $a$ field, given by 

\begin{equation}
\label{eq:OCN_D0}
J[a] =  \int_S a^{1-m} ds.
\end{equation}

Proceeding as before, but now with $b$ defined as a variation of $a$ with the boundary condition $b(x \in \Omega )= 0$, the first variation of functional \eqref{eq:OCN_D0} with the constraint of Eq. \eqref{eq:water_depth} is  
\begin{subequations}
\begin{align}
\label{eq:1st_var_1_a}
\delta J[a, b] = \frac{dJ[a + \epsilon b]}{d\epsilon}\Bigg |_{\epsilon =0}  =  & \frac{d}{d \epsilon} \int_S (a + \epsilon b)^{1-m} + \lambda \left[\nabla \cdot \left((a + \epsilon b)\frac{\nabla h}{|\nabla h|}\right) +  1 \right] ds\Bigg |_{\epsilon =0}  \nonumber \\
& =\int_S  (1-m)b(a+\epsilon b)^{-m} ds  + \int_S \lambda \left( \nabla b \cdot \frac{\nabla h}{|\nabla h|} + b  \nabla \cdot \frac{\nabla h}{|\nabla h|} \right) ds\Bigg |_{\epsilon =0} \nonumber \\
 &=\int_S  b \left[ (1-m)a^{-m}  +  \lambda \nabla \cdot \frac{\nabla h}{|\nabla h|}\right] ds  + \int_S  \lambda \nabla b \cdot \frac{\nabla h}{|\nabla h|}  ds \nonumber \\
 & =\int_S  b \left[ (1-m)a^{-m}  - \nabla \lambda  \cdot \frac{\nabla h}{|\nabla h|}\right] ds,  
\end{align}
\end{subequations}
where we used integration by parts with $b(x \in \Omega )= 0$ as the boundary condition. The first variation must be zero for any  variation $b$; therefore, we have \cite{giaquinta2013calculus1, lanczos2012variational}, 

\begin{equation}
\label{eq:1st_var_2_a}
(1-m)a^{-m}  - \nabla \lambda  \cdot \frac{\nabla h}{|\nabla h|} = 0, 
\end{equation}
which yields the sediment continuity (Eq. \eqref{eq:lem} at the steady-sate with $D=0$) for $\lambda = \frac{K(1-m)}{U} h$. 
It is interesting to note that the objective functional \eqref{eq:OCN_D0} is equivalent to functional \eqref{eq:functional_D0_C} at its critical points (refer to the section \ref{obj_D0} in the Appendix for details).

\subsection{The second variation}
We study the second variation at the critical points by writing functional \eqref{eq:functional_D0_C} as (refer to the section \ref{obj_D0} in the Appendix for details), 

\begin{equation}
\label{eq:functional_D0_C_DZ}
J[h] = \left(\frac{U}{K}\right)^{\frac{1}{m}} \int_S |\nabla h|^{\frac{m-1}{m}} ds.
\end{equation}

Defining a variation $g$ with $g(x \in \Omega) =0$ as the boundary condition, the second variation of this functional is 

\begin{subequations}
\begin{align}
\label{eq:2nd_var_D0}
\delta^2 J[h, g] = \frac{d^2J[h + \epsilon g]}{d\epsilon^2}\Bigg |_{\epsilon =0}  =  & \left(\frac{U}{K}\right)^{\frac{1}{m}} \frac{d^2}{d \epsilon^2}  \int_S |\nabla h + \epsilon \nabla g|^{\frac{m-1}{m}} ds\Bigg |_{\epsilon =0}  \nonumber \\
& =\frac{m-1}{m} \left(\frac{U}{K}\right)^{\frac{1}{m}} \frac{d}{d \epsilon}  \int_S  \nabla g \cdot  \left(\nabla h + \epsilon \nabla g\right)  \left|\nabla h + \epsilon \nabla g\right|^{-\frac{m+1}{m}}ds\Bigg |_{\epsilon =0} \nonumber \\
 &=\frac{1-m}{m^2} \left(\frac{U}{K}\right)^{\frac{1}{m}} \int_S |\nabla g|^2 |\nabla h|^{-\frac{m+1}{m}} ds
\end{align}
\end{subequations}

Thus, a critical function of $J$ is a minimum for any non-zero variation if and only if $g$ \cite{giaquinta2013calculus1} 

\begin{equation}
\label{eq:min_condition_D0}
\delta^2 J[h, g] =\frac{1-m}{m^2} \left(\frac{U}{K}\right)^{\frac{1}{m}} \int_S |\nabla g|^2 |\nabla h|^{-\frac{m+1}{m}} ds> 0. 
\end{equation}

From the constraint of the sediment continuity, we have $|\nabla h| > 0$. Given the boundary condition of the variation $g$  (i.e., $ g(x \in \Omega) =0$ ), if  $|\nabla g|=0$,  it immediately follows $g \equiv 0$. This means that the integral in RHS of Eq. \eqref{eq:min_condition_D0} is positive definite for any non-zero variation $g$. Thus, the critical functions of the variational problem in Eq. \eqref{eq:functional_D0_C} are minima for $0 < m < 1$. Following the same line of reasoning, it is easy to show that the critical functions are maxima for $m > 1$. At $m=1$, the critical functions are saddles having the same value of the objective function. This result exactly follows from integrating Eq. \eqref{eq:lem} over the domain with area $A_s$,  $D=0$, and zero elevation at the boundaries which yields $\int_{S} a |\nabla h| ds =  \frac{U}{K} A_s$. Given that $\int_{S} a |\nabla h| ds= \int_{S} h ds$ (refer to the section \ref{obj_D0} in the Appendix for details), we have  $\int_{S} h ds=\frac{U}{K} A_s$ which indicates that steady-state solutions have the same average elevation regardless of the $a$ field.

\section{Connection with OCNs}

To interpret these results in the context of the OCNs, one should notice that in a continuous domain, functional \eqref{eq:OCN_obj} is equivalent to functional \eqref{eq:OCN_D0} in which $A_i \equiv a(x,y)$ and $\gamma \equiv 1 -m$. As discussed earlier, the exponent $\gamma$ controls the structure of the OCNs. For  $0<\gamma<1$ (equivalent to $0<m<$1), each spanning tree is a local minimum \cite{banavar2000topology} for which there is a height function such that water flows toward the steepest descent \cite{balister2018river}. This is in agreement with our finding in the continuum formulation, which states the critical functions of functional \eqref{eq:functional_D0_C} should satisfy a continuity equation for $a$, constructed upon the assumption of water moving toward the surface gradient. The range $\gamma>1$ (equivalent to $m<0$) gives a configuration with the shortest path to the outlet \cite{rinaldo1996thermodynamics}, although this case is not physically possible according to the sediment continuity \cite{rinaldo2014evolution}.  

\begin{figure}[H]
  \centering
  \includegraphics[width=0.7\textwidth]{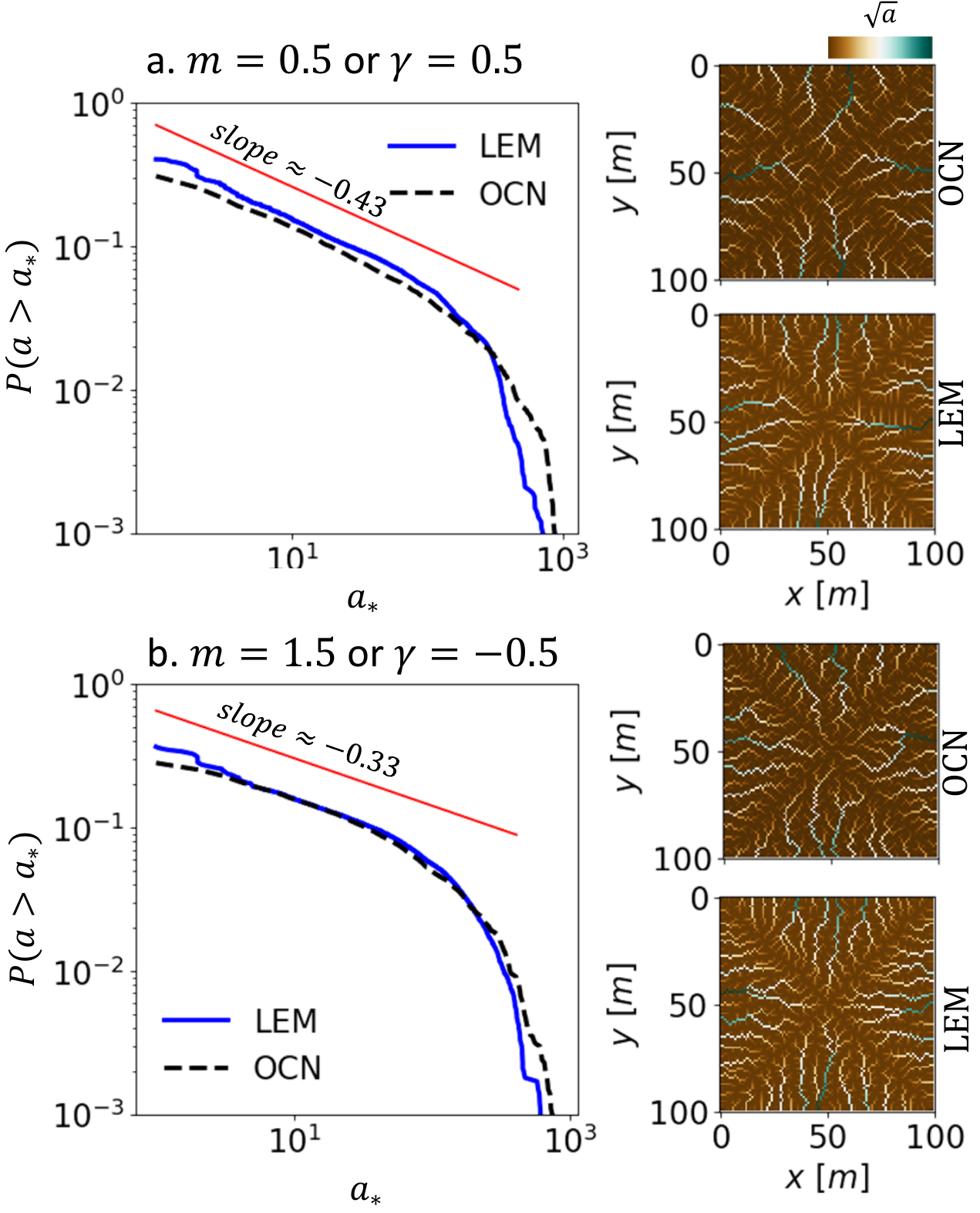}
  \caption{Distribution of the specific drainage area $a$ for (a) $m=0.5$ and (b) $m=1.5$ with a greedy optimization of Eq. \eqref{eq:OCN_D0} (denoted by OCN) and the numerical simulation of the PDEs in Eqs. \eqref{eq:lem} and \eqref{eq:water_depth} (denoted  by LEM). The fitted power functions to the first portion of the distribution from the OCN  with the scaling exponents are also shown. Both approaches are applied to the same  $100 \text{m}$ by $100 \text{m}$ domain with $1 \text{m}$ grid spacing. The resulting $a$ fields are shown in the left panels.  The optimization scheme was a greedy minimization for $m=0.5$ and maximization for $m=1.5$, in which random changes in node connectivity which led to better objective function (smaller for $m=0.5$ and larger for $m=1.5$) are accepted, until no change occurs for a given number of trials ($1000$ for these results).  } 
  \label{fig:energey_landscape}
\end{figure}

The numerical solution of the system of Eqs. \eqref{eq:lem} and \eqref{eq:water_depth} with $D=0$ (i.e., $C_I>>1$) and $m > 1$  gives  surfaces with channel networks resembling those observed in nature (see for example Fig. \ref{fig:numerical_ci}j which shows the steady-state surface for $m=1.5$ and $\mathcal{C_I} = 10^5$). On the other hand, minimizing functional \eqref{eq:OCN_obj} with $\gamma < 0$ (equivalent to $m>1$) creates loops or spiral paths as shown in Fig. \ref{fig:OCN}c. The second variation analysis performed earlier addresses this discrepancy by showing that realistic networks for OCNs with $\gamma < 0$ (i.e., $m>1$) emerge by maximizing Eq. \eqref{eq:OCN_D0} or its discretized form in Eq. \eqref{eq:OCN_obj}, whereas for $0<\gamma <1$ and $\gamma > 1$ (i.e., $0<m<1$ and $m<0$), the objective functional should be minimized.

We confirm this finding by a set of numerical experiments where we minimized Eq. \eqref{eq:OCN_D0} for $\gamma =0.5$ and maximized it for $\gamma = -0.5$ in a discretized domain following the greedy optimization algorithm explained in Section \ref{se:OCN}. In Fig. \ref{fig:energey_landscape}, the optimal $a$ fields are compared with the results from the numerical solution of Eqs. \eqref{eq:lem} and \eqref{eq:water_depth} for $m=0.5$ (i.e., $\gamma =0.5$) and $m=1.5$ (i.e., $\gamma =-0.5$). Both approaches converged to qualitatively similar $a$ fields as shown in the right panels of Fig. \ref{fig:energey_landscape}a and b. The left panels of Fig. \ref{fig:energey_landscape} show the distribution of $a$ (exceedance probability) from OCN and LEM which contain a very similar power distribution for small $a$ which is distributed by the finite-size effect for large $a$. The exponents close to  $-0.43$ as observed for $\gamma = 0.5$ (i.e., $m=0.5$) are widely reported in the literature as the signature of the feasible optimally, whereas exponents close to $-0.5$ characterize (near) global optimal state \cite{Rodriguez2001, rinaldo2014evolution}.

\section{The role of diffusion}
\begin{figure}[H]
  \centering
  \includegraphics[width=0.7\textwidth]{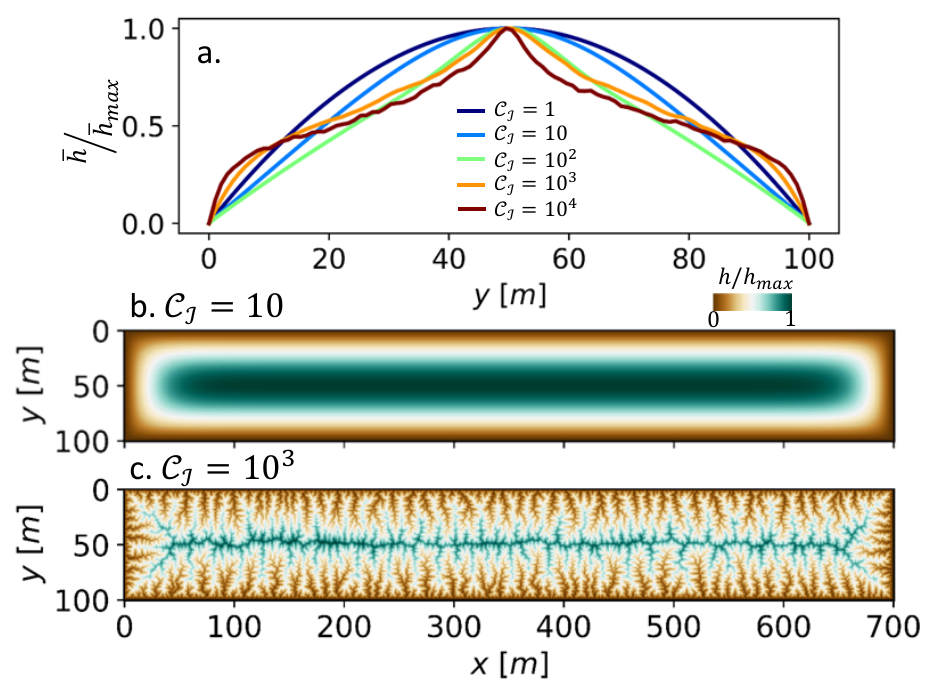}
  \caption{(a) The effect of diffusion on the mean-elevation profile, denoted by $\bar{h}$, in a long rectangular domain ($700\ \text{m}$ by $100\ \text{m}$). Higher $\mathcal{C_I}$ (relatively smaller diffusion) results in progressively more uniform mean profiles. The normalized $h$ filed for an unchannelized case with $\mathcal{C_I} = 10$ and an channelized case with $\mathcal{C_I} = 10^3$ are shown in (b) and (c). The model parameters are the same as those reported in Fig. \ref{fig:numerical_ci} and $m=0.5$.}
  \label{fig:diff}
\end{figure}

Soil diffusion transport smooths the surface and tends to prevent the formation of channels \cite{bonetti2019Cascade}. Fig. \ref{fig:diff} shows the effect of diffusive on the steady-state surfaces achieved by numerically solving the governing equations in a long rectangular domain with zero elevation at the boundaries. Smaller diffusion (higher $\mathcal{C_I}$) results in a flatter mean-elevation profile $\bar{h}$. The profiles were computed by averaging the elevation along the $x$-axis for $100<x<600 \text{m}$ to minimize the effect of the side boundaries. The choice of long domain was meant to approximate a semi-infinite case with parallel boundaries where the distance between two boundaries ($l_y = 100\ \text{m}$) is the only dominant length scale in computing $\mathcal{C_I}$. 

In this section we show that one can recover Eq. \eqref{eq:water_depth} for the general case (non-zero $K$ and $D$)  by using the sediment continuity Eq. \eqref{eq:lem} and the variational principle \eqref{eq:functional_G} (refer to the section \ref{1_var_G} in the Appendix for details).

\begin{equation}
\label{eq:functional_G}
J[h] = \int_S  h - \frac{D}{K} h \Delta a^{1 - m} ds, 
\end{equation}

It is easy to show that this functional at its critical point is equivalent to (refer to the section \ref{1_var_G} in the Appendix for details), 
\begin{equation}
\label{eq:functional_G_C_2}
J[h] = \frac{U}{K} \int_S a^{1-m} ds + \frac{D}{K} \int_\Omega  a^{1-m} (\nabla h \cdot n) d\omega , 
\end{equation}
where the second integral is a flux at the domain boundary and the first integral is the objective function derived earlier for  $D=0$, which resembles the objective functional of the OCN theory (Eq. \eqref{eq:OCN_obj}). From the numerical perspective, functional \eqref{eq:functional_G_C_2} may be used to generate locally optimal configurations; however, it needs a prior knowledge of the slope at the domain boundaries. Given the existence of diffusive transport in the formulation, one should allow the divergence of flow paths by enabling flow toward multiple nodes.   

Another a special case of this problem corresponds the condition of negligible erosion ($K=0$), in which the sediment continuity becomes the Poisson equation and decouples from the water-continuity equation, 
\begin{equation}
\label{eq:lem_K0}
\frac{\partial h}{\partial t}=D\Delta h + U. 
\end{equation}

At steady-state, the solution of this equation corresponds to the critical function of the functional (refer to the section \ref{var_NE} in the Appendix for details), 

\begin{equation}
\label{eq:functional_K0}
J[h] = \int_S h + \frac{D}{2U} |\nabla h|^2 ds.  
\end{equation}

This functional without $h$ in the integrand is the Dirichlet integral and is similar to the variational formulation of  the Fokker–Planck equation associated with the Brownian motion \cite{courant1928partiellen, jordan1998variational}. By analyzing the second variation, it is easy to show that the critical function of this functional is a minimum for $\frac{D}{U} > 0$ (refer to the section \ref{var_NE} in the Appendix for details). 

\section{Conclusion}
The minimalist LEM considered here has been quite successful in capturing the essential dynamics of surface evolution to reproduce statistically similar surfaces to those observed in nature \cite{perron2009formation, hancock2002testing}. On the other hand, networks generated by minimizing the total energy dissipation have been shown to embody several scaling laws observed in natural river networks \cite{Rodriguez2001}, hinting at a possible connection between these two fundamental theories of  landscape evolution \cite{banavar2000topology, banavar2001scaling, rinaldo2014evolution, balister2018river}.
We have showed here that each equation of this LEM can be independently replaced by a variational principle in the absence of the diffusive transport, a finding that proves the tendency of landscapes toward an optimal state. 

Interestingly, the properties of the optimal state (critical functions) vary based on a model parameter $m$ (exponent of the specific drainage area in the erosion term). For $0<m<1$, the optimal states are minima, for $m>1$ they are maxima, and at $m=1$, which corresponds to a saddle point, the objective functional is independent of the surface connectivity and the specific drainage area. This result is relevant for the OCN theory: OCNs are generally associated with a specific range of the parameter $\gamma$ (i.e., $0<\gamma<1$, where $\gamma \equiv 1-m$). However, our results suggest that OCN theory applies to $\gamma < 0$ where the objective functional should be maximized.

% \enlargethispage{20pt}

% \ethics{Insert ethics text here.}

% \dataccess{Insert data access text here.}

% \aucontribute{Insert author contribute text here.}

% \competing{Insert competing text here.}

% \funding{Insert funding text here.}

% \ack{Insert acknowledgment text here.}

% \disclaimer{Insert disclaimer text here.}

\section{Appendix}

\subsection{Objective functional in the absence of diffusion}\label{obj_D0}

In this section we show that functional \eqref{eq:functional_D0_C} at its critical points is equivalent to functional \eqref{eq:OCN_D0}. We begin by substituting $\lambda = \frac{1}{K}a^{1-m}$ in Eq. \eqref{eq:functional_D0_C},

\begin{equation}
\label{eq:functional_D0_C_a}
J[h] = \int_S \left(h - a |\nabla h| + \frac{U}{K}a^{1-m} \right) ds, 
\end{equation}

\begin{figure}[H]
  \centering
  \includegraphics[width=0.7\textwidth]{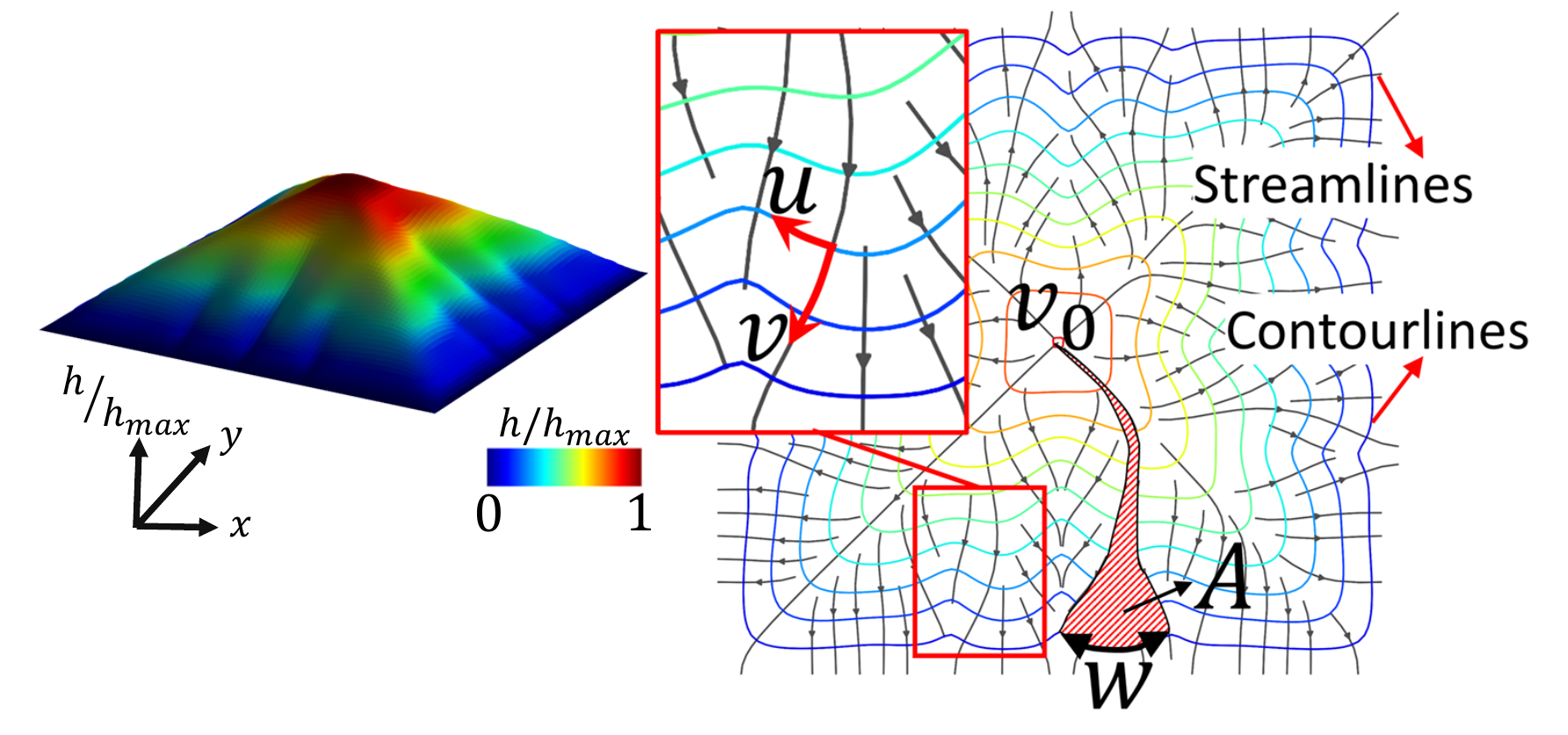}
  \caption{Curvilinear coordinate system $(u, v)$, where $v$ is in the direction of the surface gradient and is in the direction of streamlines, whereas $u$ is perpendicular to $v$ and in the direction of contourlines. The schematic definition of contour length $w$ and catchment area $A$ in Eq. \eqref{eq:a_curve} are also shown. }
  \label{fig:coordinates}
\end{figure}

To compute the term $\int_S a |\nabla h| ds$, it is convenient to map the integral into a  curvilinear coordinate system $(u, v)$, where $v$ is in the direction of the surface gradient and $u$ is perpendicular to $v$ (Fig. \ref{fig:coordinates}). Given that the streamlines are defined in the direction of the gradient, the coordinates $v$ and $u$ are tangent to the streamlines and contour lines, respectively. In such an orthogonal curvilinear coordinates, the element of area is defined as $ds = \mathcal{J} du dv$, where $\mathcal{J}$ is the Jacobian defined as \cite{stoker1989differential, gallant2011differential} 

\begin{equation}
\label{eq:jac}
\mathcal{J} = \left|x_u y_v - x_v y_u\right|,
\end{equation}
where the subscripts denote the derivative with respect to that coordinate. The elements of length along $u$ and $v$ coordinates are also defined as $dw=\sqrt{E} du$ and $dl= \sqrt{G} dv$, where  $E = x_u^2 + y_u^2$, and $G  = x_v^2 + y_v^2$. Given the orthogonality of $u$ and $v$, we have $\mathcal{J}=\sqrt{EG}$ \cite{stoker1989differential, gallant2011differential}. 

The specific catchment area $a$ from Eq. \eqref{eq:water_depth} is by definition the catchment area $A$ per unit contour length $w$ for $w \to 0$ \cite{bonetti2018theory}, which can be transferred to curvilinear coordinates through the following integral equations \cite{gallant2011differential}

\begin{equation}
\label{eq:a_curve}
 a = \lim_{w \to 0} \frac{A}{w}  =\frac{1}{\sqrt{E}}\int_v \mathcal{J} dv. 
\end{equation}

Using $a$  in Eq. \eqref{eq:a_curve} and given that $|\nabla h| =-\frac{\partial h}{\partial l}$,  where $dl=\sqrt{G} dv$ is the length of element along $v$ (i.e., along the streamline), the integral $\int_S h a |\nabla h| ds$  is

\begin{equation}
\label{eq:int_flov_1}
\int_S a |\nabla h| ds = \int_{u}\int_{v} \left(\frac{1}{\sqrt{E}} \int_{v} \mathcal{J} dq\right)\frac{1}{\sqrt{G}} \frac{\partial h}{\partial v} \mathcal{J} dv\ du. 
 \end{equation}
 
Integrating by parts gives 

\begin{equation}
\label{eq:int_flov_2}
\int_S  a |\nabla h| ds = \int_u \left[ h  \left( \int_{v} \mathcal{J} dq\right) \Bigg |_0^{v_{t}} - \int_v h \frac{\partial}{\partial v}\left( \int_{v}\mathcal{J} dq\right) dv  \right] du.  
 \end{equation}

At the initiation point of each streamline ($v=0$), we have $\int_{v} \mathcal{J} dq =0$.  The terminal point ($v=v_{t}$) where a streamline flows out of the domain is in fact a point along the boundaries in our case, where we have already enforced $h=0$ as the boundary condition; therefore, 

\begin{equation}
\label{eq:int_flov_first_term}
h  \left( \int_{v} \mathcal{J} dq\right) \Bigg |_0^{v_{t}} =0.
 \end{equation}

The derivative with respect to $v$ in the second term of the integral in Eq. \eqref{eq:int_flov_2} is simplified as 

\begin{equation}
\label{eq:int_flov_3}
\int_S a |\nabla h| ds= \int_u \int_v h \mathcal{J} dv du = \int_S h ds.   
 \end{equation}
 
Thus, Eq. \eqref{eq:functional_D0_C_a} is simplified as \eqref{eq:OCN_D0}, from which it is easy to write the functional in term of gradient (Eq. \eqref{eq:functional_D0_C_DZ}) using the sediment balance equation.

\subsection{First variation in the general case}\label{1_var_G}
Adding the steady-state sediment continuity (Eq. \eqref{eq:lem}) to functional \eqref{eq:functional_G} as the constraint with the Lagrange multiplier field $\lambda$ gives 

\begin{equation}
\label{eq:functional_G_C}
J[h] = \int_S  h - \frac{D}{K} h \Delta a^{1 - m} + \lambda \left(D\Delta h-K a^m |\nabla h| + U \right) ds. 
\end{equation}

Following similar steps as for the case $D=0$  and using an additional boundary condition on the variation $g$ (i.e. $\nabla g \cdot n \Bigg |_{\Omega}  = 0$ ), the first variation of this functional is 

\begin{equation}
\label{eq:euler_lagrange_1}
\left(1 - \frac{D}{K} \Delta a^{1 - m}\right) + K \nabla \cdot \left(\lambda a^m \frac{\nabla z}{|\nabla z|}\right) + D \Delta \lambda = 0,
\end{equation}
which is equivalent to the water balance equation (Eq. \eqref{eq:water_depth}) with  $\lambda = \frac{1}{K}a^{1-m}$. At a stable point of Eq. \eqref{eq:functional_G_C},  we have  $\lambda = \frac{1}{K}a^{1-m}$  and the water and sediment continuity equation also hold; therefore, Eq. \eqref{eq:int_flov_3} is valid and Eq. \eqref{eq:functional_G_C} is  

\begin{equation}
\label{eq:functional_G_C_1}
J[h] = \frac{D}{K} \int_S  a^{1-m} \Delta h ds - \frac{D}{K} \int_S h \Delta a^{1 - m} ds + \frac{U}{K} \int_S a^{1-m} ds. 
\end{equation}

Integration by parts of the first two terms and using the boundary condition ($h_{\Omega} = 0$) leads to \eqref{eq:functional_G_C_2}. 

\subsection{Variations in the absence of erosion}\label{var_NE}

Similarly to the case $D=0$ and with the same definition for $g$, the first variation is defined as 

\begin{subequations}
\begin{align}
\label{eq:1st_var_1_K0}
\delta J[h, g] = \frac{dJ[h + \epsilon g]}{d\epsilon}\Bigg |_{\epsilon =0}  =  & \frac{d}{d \epsilon} \int_S (h + \epsilon g) + \frac{D}{2U} |\nabla h + \epsilon \nabla g|^2 ds\Bigg |_{\epsilon =0} \nonumber \\
& =\int_S  g  + \frac{D}{U}   \nabla g \cdot (\nabla h + \epsilon \nabla g)   ds\Bigg |_{\epsilon =0} \nonumber \\
 &=\int_S  g  + \frac{D}{U}   \nabla g \cdot \nabla h  ds \nonumber \\
 &=\int_S  g \left(1   + \frac{D}{U}  \Delta h\right)  ds
\end{align}
\end{subequations}

while the second variation is 

\begin{subequations}
\begin{align}
\label{eq:2st_var_1_K0}
\delta^2 J[h, g] = \frac{d^2J[h + \epsilon g]}{d\epsilon^2}\Bigg |_{\epsilon =0}  =  & \frac{d^2}{d \epsilon^2} \int_S (h + \epsilon g) + \frac{D}{2U} |\nabla h + \epsilon \nabla g|^2 ds\Bigg |_{\epsilon =0}  \nonumber \\
& = \frac{D}{U}  \int_S |\nabla g|^2 ds 
\end{align}
\end{subequations}

%%%%%%%%%% Insert bibliography here %%%%%%%%%%%%%%
\bibliographystyle{unsrt}
\bibliography{ref.bib}

\end{document}